\documentclass[twocolumn,prl,a4paper,superscriptaddress]{revtex4}
\usepackage{comment}
\usepackage{color}
\usepackage{amsmath}
\usepackage{amssymb}
\usepackage{graphicx}
\usepackage{braket} 
\usepackage{esint}
\usepackage{adjustbox} 
\usepackage[utf8]{inputenc}

\usepackage[unicode=true,pdfusetitle,
bookmarks=true,bookmarksnumbered=false,bookmarksopen=false,
breaklinks=true,pdfborder={0 0 1},backref=false,colorlinks=true]
{hyperref}
\hypersetup{
	linkcolor=red,urlcolor=blue,citecolor=blue,anchorcolor=blue} 
\usepackage{ulem} 

\usepackage{dcolumn}\usepackage{bm}

\makeatother

\begin{document}

\title{Acute entanglement and Photon/Phonons statistics in a balanced/unbalanced $PT$-symmetry systems  }
\author{Muhammad Abid} 
\author{Areeda Ayoub}   
\author{Javed Akram}
\email{javedakram@daad-alumni.de}

\affiliation{Department of Physics, COMSATS University Islamabad, Pakistan}

 \date{\today}
\begin{abstract}

We study the significance of Photon/Phonons bunching and antibunching on the dynamics of the quantum entanglement in the presence of coupled $PT$-symmetry systems with balanced/unbalanced gain and loss. We suggest a hybrid electromechanical system to realize a strong and tunable coupling between a Coplanar-Waveguide (CPW) microwave cavity and a nanomechanical resonator  (NAMR) via a superconducting Transmon qubit.  The hybrid electromechanical system consists of a non-hermitian Hamiltonian with balanced/unbalanced gain and loss.  The interplay between the quantum entanglement and the $PT$-symmetry systems is also thoroughly investigated.  We frame a connection between Number operators, Photon/Phonons antibunching, and entanglement. It has been observed that the relative Photon/Phonons numbers play a key role in the quantum entanglement dynamics. Furthermore,  we study that quantum entanglement can be characterized by defining a Photon/Phonons antibunching. The Photon/Phonons antibunching is strongly dependent on the initial squeezed state and the rate of balanced/unbalanced gain and loss of the system. 
	
\end{abstract} 

\pacs{42.50.Nn, 42.50.Dv, 03.65.−w, 03.70.+k   }  
		  
\maketitle

\section{Introduction}  \label{Sec-1}
 Recently, the most promising idea pursued by the researchers is to combine two or more  systems to form hybrid quantum systems, intending to harness the strengths and advantages of different physical systems in order to explore new physical phenomena and potentially give rise to novel quantum technologies \cite{RevModPhys.75.457,Hanson-2008,PhysRevB.77.014510,Cheng-2015,PhysRevB.72.195411,PhysRevA.76.024301,Xue-2007,Akram-2008,RevModPhys.85.623,PhysRevB.76.064305,PhysRevB.87.014513,PhysRevLett.104.177205,PhysRevLett.95.097204,PhysRevLett.103.227203,PhysRevA.79.052102,PhysRevA.82.032101,Pirkkalainen-2015,HUSSAIN2020412152}. They provide new tools and platforms to investigate deeper into unexplored quantum regimes. Among these systems, the nanomechanical resonators  (NAMRs) embedded in qubits (including artificial  atoms \cite{RevModPhys.75.457}, spins \cite{Hanson-2008}, superconducting qubits \cite{PhysRevB.77.014510,Cheng-2015}, and so on) are considered as a good candidate for exploring different quantum phenomena. For example, quantum entanglement \cite{PhysRevB.72.195411,PhysRevA.76.024301,AYOUB20211353977}, controllable coupling \cite{Xue-2007}, hybrid quantum circuits \cite{RevModPhys.85.623}, quantum squeezing \cite{PhysRevB.87.014513,PhysRevB.76.064305}, quantum detection in nanomechanical systems \cite{PhysRevLett.104.177205}, ground-state cooling \cite{PhysRevLett.95.097204,PhysRevLett.103.227203,PhysRevA.79.052102} and phonon blockade \cite{PhysRevA.82.032101} have already been proposed. Moreover, a controllable coupling between NAMR and Coplanar-Waveguide (CPW) microwave cavity has been realized through a qubit \cite{Pirkkalainen-2015}.
 
 On the other hand, the $PT$-symmetry has been studied in different quantum  systems \cite{Ruter-2010,PhysRevA.86.013612,PhysRevLett.113.053604,PhysRevA.92.013852,PhysRevLett.110.083604,PhysRevA.89.023601,PhysRevA.97.053846,Akram:21}. A wide class of non-hermitian Hamiltonians respecting $PT$-symmetry exhibit entirely real spectrum of eigenvalues. The most interesting feature of such Hamiltonians is $PT$ phase transition, in which the eigenvalues spectrum switches from being entirely real to being completely imaginary, which is marked by the presence of exceptional point (EP), where two or more eigenvalues and their corresponding eigenvectors coalesce and become degenerate \cite{Ruter-2010}. Therefore, many interesting physical properties can also be studied with the $PT$ symmetric device the dynamics of which is governed by an intrinsic quantum-mechanical law. For example, PT-symmetric  Bose-Einstein condensate in a $\delta$-function double-well potential  \cite{PhysRevA.86.013612,Hussain-2019}, optomechanical devices as a phonon laser \cite{PhysRevLett.113.053604,PhysRevA.92.013852}, non-linear dynamics in cold atoms \cite{PhysRevA.89.023601}, and $PT$- symmetric circuit QED systems \cite{PhysRevA.97.053846} have been proposed. Many intrinsic quantum properties have been studied with the help of $PT$ devices such as decoherence dynamics \cite{PhysRevA.94.040101}, information retrieval and criticality \cite{PhysRevLett.119.190401}, entanglement \cite{PhysRevA.90.054301}, and chiral population transfer \cite{Xu-2016,Doppler-2016}. Moreover, due to the considerable progress in theories and experiments for the generation of mechanical gain and cavity loss \cite{Zhang-2018, PhysRevA.97.053830}, more interesting properties have been studied. However, the true quantumness of $PT$ system is still questionable because to preserve the proper commutation relation, quantum noises associated with the gain (amplifying) and loss (dampening) are often ignored, which shows a drastic difference than usual predictions \cite{PhysRevA.85.031802,Kepesidis-2016,PhysRevLett.123.180501}. 
 
 Entanglement, being an inherent form of quantum correlation, has become an invaluable resource for quantum computing and quantum information processing. As it is well known that quantum entanglement is very fragile under the influence of environmental noises \cite{PhysRevLett.93.140404,Huang-2008}. Notably, due to the presence of quantum noises, the continuous variable (CV) entanglement generated in a system of two coupled waveguides is badly affected \cite{PhysRevA.96.033806}.  Therefore, experimentalists are mostly concerned to preserve the entanglement for a long time, as unavoidable interaction with an external environment is significantly detrimental to the generation of entanglement. A wide variety of decoherence entangled pairs such as photons \cite{PhysRevLett.99.180504}, atoms \cite{Almeida-2007}, continuous Gaussian states \cite{PhysRevLett.100.220401}, and spin chains \cite{PhysRevA.78.012357} have been analyzed both experimentally and theoretically for the environment-induced sudden death of entanglement. The presence of 
quantum noise enables us to understand their effects in broken $PT$ -symmetric systems more deeply.
  
 In this article, we present a general framework for two paired quantum systems that adhere to $PT$-symmetry. Relevant linear quantum systems shared balanced/unbalanced gain and loss with their neighbors. By defining unbalanced gain and loss, we try to understand the effects of quantum noises on entanglement. In this work, we describe the dynamical behavior of entanglement at exceptional points in the presence of balanced/unbalanced gain and loss. We study the effect of $PT$-symmetry exceptional point on the evolution of the number of photons and phonons with balanced/unbalanced gain and loss. We investigate the dynamics of a quantum correlation function in the different regimes across the exceptional points balanced/unbalanced gain and loss and try to build a relation between photon/phonon numbers and phase coherence. 
The rest of the paper is organized as follows. In Sec. \ref{Sec-2}
we present the description of the hybrid electromechanical system with the non-hermitian Hamiltonian. A hybrid electromechanical system to realize a strong and tunable coupling between a CPW microwave cavity and a NAMR via a superconducting Transmon qubit. The direct coupling between
NAMR and CPW microwave cavities would be weak and
uncontrollable as well due to their size mismatch. 
In Sec. \ref{Sec-3}, we determine the effective Hamiltonian between the  CPW microcavity and the NAMR by using the Fröhlich-Nakajima transformation. In Sec. \ref{Sec-4}, we illustrate the evolution of entanglement, photon/phonon numbers, and a kind of quantum correlation function in different regimes across the exceptional points. In the same section, we describe the effect of quantum noise on the dynamics of entanglement and the growth of photon/phonon numbers. Additionally, we also study the influence of squeezing on the dynamics of entanglement. In Sec. \ref{Sec-5}, we discuss in detail how to do altering of the entanglement dynamics by changing the initial squeezing parameter for balanced and unbalanced scenarios.  Finally, we conclude our findings in Sec. \ref{Sec-6}.

\section{Hamiltonian of the hybrid system }  \label{Sec-2}
\begin{figure}
    \centering
    \includegraphics[height=8cm,width=8cm]{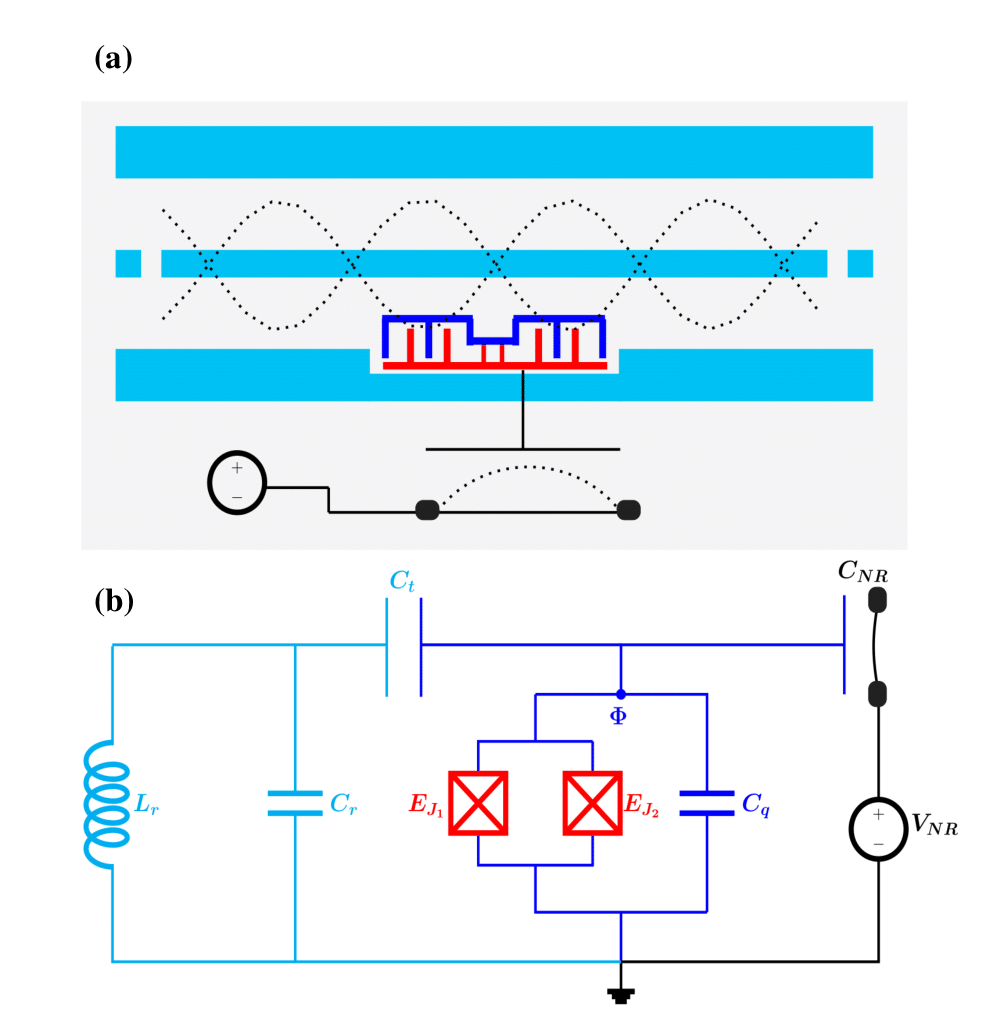}
\caption{(a) Schematic description of the hybrid electromechanical quantum system, which comprises a CPW microwave cavity, a superconducting qubit, and a NAMR. The coupling between the Transmon qubit and the NAMR can be achieved by applying external dc voltage in the NAMR. (b) Equivalent circuit of the hybrid quantum system.}
    \label{Fig-1}
\end{figure}

We consider a hybrid electromechanical system, which consists of a superconducting CPW microwave cavity, a superconducting Transmon qubit, and a NAMR, as depicted in Fig. (\ref{Fig-1}). The Hamiltonian of CPW microwave cavity can be expressed as $H_c=\hbar\omega_c c^{\dag}c$, where $c (c^{\dag})$ denotes the bosonic annihilation (creation) operator of the cavity mode with resonant frequency $\omega_c=\sqrt{1/L_{r}C_{r}}$. Here, $L_r$ and $C_r$, respectively, represent the total inductance and capacitance of the cavity. The Hamiltonian of the NAMR is defined as $H_m=\hbar\omega_m b^{\dag}b$, with $b(b^{\dag})$ being the bosonic annihilation (creation) operator of the mechanical mode and $\omega_m$ is the resonant frequency of the NAMR. The Hamiltonian of superconducting Transmon qubit is given as $H_q=4E_{C}(n-n_g)^2-E_{J}\cos{\phi}$ \cite{PhysRevA.76.042319}, where $n$ denotes the number of Cooper pairs, $n_g$ describes the effective offset charge due to environmental sources and $E_{C}=e^{2}/2C_{eq}$ defines the charging energy of the capacitor, in which $C_{eq}=C_{NR}(x)+C_t+C_q$ is the total capacitance. $C_{NR}(x)=C_{NR}(1-x/d)$ describes the capacitive coupling between the Transmon qubit and the NAMR as a function of resonator displacement $x$. Normally the length $d$ is large as compared to the oscillation displacement $x$. 
  $C_t$ illustrates the coupling capacitance between the Transmon qubit and the CPW microwave cavity and $C_{q}$ defines the total capacitance of the Transmon qubit.   The effective Josephson coupling energy defines the Transmon qubit is given as $E_J=E_{J_{1}}+E_{J_{2}}=E_{J_{0}}\cos{(\pi\Phi_{ext}/\Phi_0})$, where $E_{J_{0}}$ is the maximum Josephson energy, $\Phi_{ext}$ is the externally applied magnetic flux, and $\Phi_{0}$ is the superconducting magnetic flux quantum. In the rotating wave approximation, the effective Hamiltonian of the coupled system reads ($\hbar=1$)\cite{PhysRevLett.114.173602}

\begin{equation}\label{a1}
    H=\omega_c c^{\dag}c+\omega_m b^{\dag}b+\frac{\omega_q}{2}\sigma_z+g\Big(c^{\dag}\sigma_{-}+c\sigma_{+}\Big)+\lambda\Big(b^{\dag}\sigma_{-}+b\sigma_{+}\Big),
\end{equation}
 where $\omega_q$ and $\sigma_z=\ket{e}\bra{e}-\ket{g}\bra{g}$ denote, respectively, the transition frequency and the  Pauli operator for the Transmon qubit. $\sigma_{+}=\ket{e}\bra{g}(\sigma_{-}=\ket{g}\bra{e})$ is the raising (lowering) operator, where $\ket{e}(\ket{g})$ describes the excited (ground) state of the Transmon qubit. The coupling strength between the CPW microwave cavity and the Transmon qubit is defined as $g=-2{V_{rms}^{0}}C_t/C_{eq}\hbar$ , where $V_{rms}^{0}=\sqrt{\hbar\omega_{c}/2C_{r}}$ is the \textit{rms} of vacuum voltage fluctuations of the CPW microwave cavity. $\lambda=x_{zpf}(4E_{c}V_{NR}/e\hbar)dC_{NR}/dx$ is the coupling strength between the NAMR and the qubit, where $x_{zpf}$ represents the zero-point fluctuation of the NAMR oscillation displacement \cite{PhysRevB.68.155311, RevModPhys.86.1391}, and $V_{NR}$ is the applied external dc voltage to the NAMR. Thus, the physical properties of the hybrid quantum system can be adjusted by changing the external voltage, which modulate the coupling $\lambda$ between the qubit and the NAMR.

 \section{ Effective Coupling between The CPW microwave cavity and the NAMR }\label{Sec-3}
 
 To decouple the Transmon qubit from hybrid quantum system, we use Fr\"{o}hlich-Nakajima transformation \cite{PhysRev.79.845} to Hamiltonian
 \begin{equation}\label{a2}
     H_{eff}=e^{-S}He^{S}=H+[H,S]+[[H,S],S]/2+...,
 \end{equation}
 where
 \begin{equation}\label{a3}
     S=\frac{g}{\Delta_c}\Big(c^{\dag}\sigma_{-}+c\sigma_{+}\Big)+\frac{\lambda}{\Delta_m}\Big(b^{\dag}\sigma_{-}+b\sigma_{+}\Big).
 \end{equation}
 We have considered the hybrid quantum system in dispersive regime where the detuning frequencies of the Transmon qubit from the CPW microwave cavity ($\Delta_c\equiv\omega_{q}-\omega_{c}>0$) and NAMR ($\Delta_m\equiv\omega_{q}-\omega_{cm}>0$), respectively, are much larger than the coupling strengths $g$ and $\lambda$ i.e., $g/\Delta_c<<1$ and $\lambda/\Delta_m<<1$. Since, both the coefficients $g/\Delta_{c}$ and $g/\Delta_{m}$ are very small, the higher-order terms can be neglected and only the second-order term needs to be kept in Eq. (\ref{a2}). Therefore, the effective Hamiltonian can be rewritten as
 \begin{equation}\label{a4}
     H_{eff} \approx \omega_{c0}c^{\dag}c+\omega_{m0}b^{\dag}b-G(c^{\dag}b+cb^{\dag}),
 \end{equation}
 where the effective microwave cavity frequency $\omega_{c0}$,  the effective NAMR frequency $\omega_{m0}$, and the effective coupling strength $G$ between the microwave cavity and the NAMR are defined as
 \begin{equation*}
     \omega_{c0}=\omega_{c}-\frac{g^2}{\Delta_c},\hspace{0.5cm} \omega_{m0}=\omega_{m}-\frac{\lambda^2}{\Delta_m},
 \end{equation*}
 \begin{equation}\label{a5}
      G=g\lambda\Big(\frac{\Delta_{c}+\Delta_{m}}{2\Delta_{c}\Delta_{m}}\Big).
 \end{equation}
While deriving Eq. (\ref{a4}), we have assumed that the Transmon qubit is at the ground state  ($\sigma_{+}\sigma_{-}=0$). In this way, an effective coupling between the CPW microwave cavity and NAMR can be achieved by adiabatically eliminating the degrees of freedom of the qubit. From Eq. (\ref{a5}), we see that the frequencies of both the cavity mode and the mechanical mode are shifted due to their large detunings with the qubit, and the effective electromechanical coupling strength $G$ depends upon the coupling strengths, $g$ and $\lambda$, and the frequency detunings, $\Delta_c$ and $\Delta_m$. By modulating the NAMR-qubit coupling strength $\lambda$ with the external voltage, the effective electromechanical coupling $G$ can be controlled.

\section{Entanglement in \textit{PT}-Symmetric electromechanical system}\label{Sec-4}

By considering the effect of the thermal environment,  the  quantum Heisenberg-Langevin equations for the system
are written as
\begin{subequations}\label{a6}
\begin{align}
\Dot{c}=-\Big(\Dot{\iota}\omega_{c0}+\frac{\kappa}{2}\Big)c+\Dot{\iota}Gb+\sqrt{\kappa}c^{in},\label{6a} \\
\Dot{b}=-\Big(\Dot{\iota}\omega_{m0}-\frac{\gamma}{2}\Big)b+\Dot{\iota}Gc+\sqrt{\gamma} b^{in},\label{6b} 
\end{align}
\end{subequations}
where $\kappa$ defines the cavity loss and $\gamma$ describes the mechanical gain. The vacuum input noise operators are defined as $c^{in}$ and $b^{in}$, which satisfying following correlation functions:
\begin{subequations}\label{a7}
\begin{align} 
\langle c^{in \dag} (t){c^{in}}(t')\rangle=0,\label{7a}\\
\langle c^{in}(t){c^{in \dag}} (t')\rangle=\delta(t-t'),\label{7b} \\
\langle b^{in \dag} (t){b^{in}}(t')\rangle=n_{th}\delta(t-t'),\label{7c} \\
\langle b^{in}(t){b^{in\dag}} (t')\rangle=(n_{th}+1)\delta(t-t'),\label{7d}
\end{align}
\end{subequations}
where $n_{th}=[exp(\hbar \omega_{m0}/k_{B}T)-1]^{-1}$ illustrates the mean thermal photon numbers of the mechanical resonator at temperature $T$ and $k_B$ describes the Boltzmann constant. Under the Markovian assumption, the noise operators $c^{in}$ and $b^{in}$ have zero mean values. Next, by introducing two slowly varying operators, $\Tilde{c}=c e^{\Dot{\iota}\omega_{c0}t}$ and $\Tilde{b}=be^{\Dot{\iota}\omega_{m0}t}$, Eq. (\ref{a6}) can be rewritten as
\begin{subequations}\label{a8}
\begin{align}
\Dot{\Tilde{c}}=-\frac{\kappa}{2}\Tilde{c}+\Dot{\iota}G\Tilde{b}e^{\Dot{\iota}(\omega_{c0}-\omega_{m0})t}+\sqrt{\kappa}\Tilde{c}^{in},\label{8a} \\
\Dot{\Tilde{b}}=\frac{\gamma}{2}\Tilde{b}+\Dot{\iota}G\Tilde{c}e^{-\Dot{\iota}(\omega_{c0}-\omega_{m0})t}+\sqrt{\gamma}\Tilde{b}^{in}.\label{8b}
\end{align}
\end{subequations}
Here, we define two  noise operators $\Tilde{c}^{in}=c e^{\Dot{\iota}\omega_{c0}t}$ and $\Tilde{b}^{in}=b e^{\Dot{\iota}\omega_{m0}t}$, which possess the  correlation functions as described in Eq. (\ref{a7}). We assume that the CPW microwave cavity is resonant with NAMR i.e., $\omega_{c0}=\omega_{m0}$ to obtain
    \begin{subequations}\label{a9}
\begin{align}
\Dot{\Tilde{c}}=-\frac{\kappa}{2}\Tilde{c}+\Dot{\iota}G\Tilde{b}+\sqrt{\kappa}\Tilde{c}^{in},\label{9a} \\
\Dot{\Tilde{b}}=\frac{\gamma}{2}\Tilde{b}+\Dot{\iota}G\Tilde{c}+\sqrt{\gamma}\Tilde{b}^{in}.\label{9b}
\end{align}
\end{subequations}
Next, we ignore quantum noises to recast Eq. (\ref{a9}) as $\Dot{u}(t)=-\Dot{\iota}\Tilde{H}u(t)$. Here, $u^{T}(t)=(q_{1}(t),p_{1}(t),q_{2}(t),p_{2}(t))$ is the state vector, which can be written in terms of dimensionless CV quadrature as $q_{1}\equiv(c+c^{\dag})/\sqrt{2}$, $p_{1}\equiv(c-c^{\dag})/\Dot{\iota}\sqrt{2}$, $q_{2}\equiv(b+b^{\dag})/\sqrt{2}$ and $p_{2}\equiv(b-b^{\dag})/\Dot{\iota}\sqrt{2}$. The non-Hermitian Hamiltonian of our system can be defined as

\begin{equation}\label{a10}
    \Tilde{H} = \Dot{\iota}\begin{pmatrix} 
   \frac{\gamma}{2}&  0 & 0&-G \\\\
  0 &  \frac{\gamma}{2} & G&0\\\\ 0&-G&-\frac{\kappa}{2}&0\\\\G&0&0&-\frac{\kappa}{2}
    \end{pmatrix}.
 \end{equation}
 
It can be easily verified that the Hamiltonian $ \Tilde{H}$ remains invariant under the simultaneous $PT$  operation i.e., $[PT, \Tilde{H}]=0$.  To study the $PT$ phase transition, we first diagonalize the Hamiltonian and find the eigen frequencies of two supermodes
\begin{equation}\label{a11}
\omega_{\pm}=\frac{\dot{\iota}(\gamma-\kappa)}{4} \pm\sqrt {G^2 -\left(   \frac{\gamma+\kappa }{4} \right )^2}
\end{equation}
where the real and imaginary parts correspond to the effective frequency and the dissipation in the system, respectively. From Eq. (\ref{a11}), we find that the  eigenfrequencies $\omega_{\pm}$ of the two modes are highly dependent on the coupling strength $G$ and there is a critical point $G_{c}=(\gamma+\kappa)/4$. Next, we consider the general situation where the mechanical gain and the cavity loss are defined with respect to each other  i.e., $\gamma=s\kappa$, where $s$ defines the ratio of gain and loss. By defining gain in terms of loss, we can rewrite critical coupling strength as   $G_c = \kappa(s + 1)/4$.

\begin{figure}
    \centering
  \includegraphics[height=3.8cm,width=8.8cm]{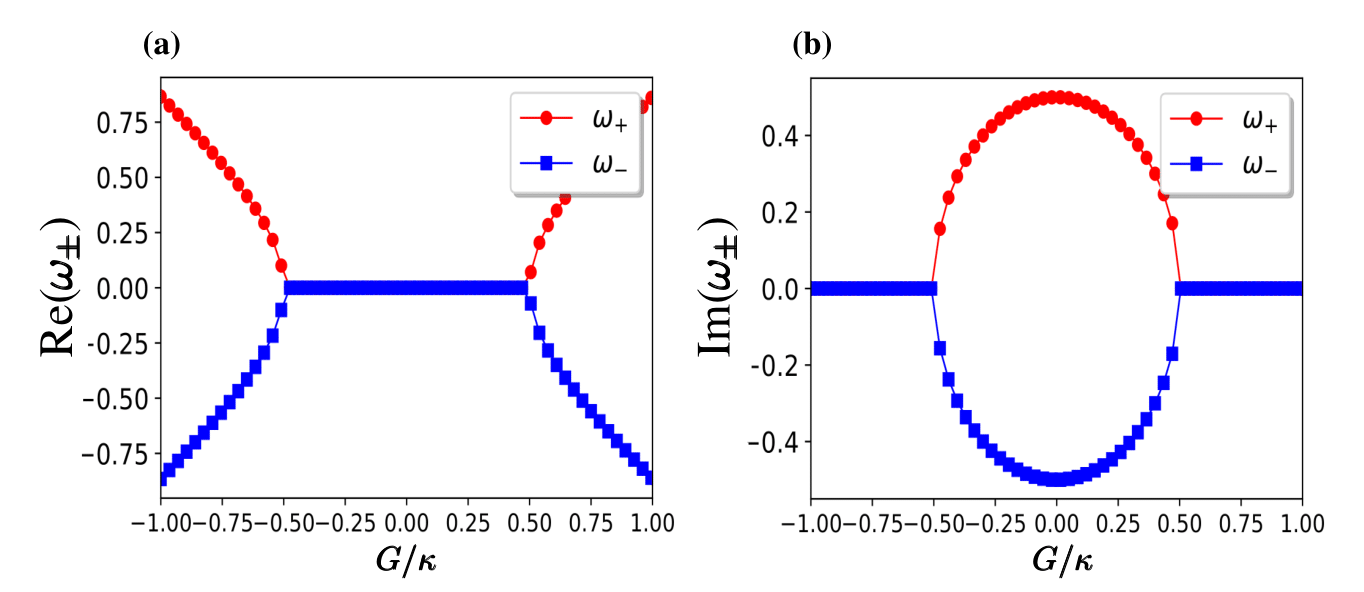}
\caption{For the case of balanced gain-loss i.e., $\gamma=\kappa$ (a) the real and (b) the imaginary  parts the eigen frequencies $\omega_{\pm}$ are plotted as a function of dimensionless coupling strength $G/\kappa$.}
    \label{Fig-2}
\end{figure}

\subsection{Balanced Gain $\&$ Loss ($s=1$)}
In this subsection,  we first study the dynamics of the hybrid electromechanical system for the case when the cavity loss and the mechanical gain are strictly balanced i.e., $\gamma=\kappa (s=1)$. In this situation, the value of the critical coupling strength becomes $|G_{c}|=\kappa/2$. For the case of balanced gain and loss, the real and the imaginary parts of the eigenfrequencies $\omega_{\pm}$ are plotted as a function of dimensionless coupling strength $G/\kappa$, as shown in Fig. \ref{Fig-2}(a, b). It can be seen in Fig. \ref{Fig-2}(a, b), for $|G|>\kappa/2$ and $|G|<\kappa/2$, respectively, there exist two distinct phases: one that includes all the real eigenvalues named as the $PT$ symmetric phase and the other which possesses purely imaginary eigenvalues named as broken $PT$ symmetric phase. Both of these phases are separated by critical coupling strength $|G_{c}|=\kappa/2$, also known as an exceptional point (EP).
 Physically, the EP refers to the situation where the mechanical coupling rate $G$ is balanced with the respective mechanical heating and cooling rate. Then, for $|G|> {\kappa}/{2} $, we get a coherent exchange of energy between the CPW microwave cavity and the NAMR which corresponds to the $PT$ symmetric phase and for $|G|< {\kappa}/{2}$  i.e., the coupling strength becomes weak enough to support the energy exchange, there is a localization of energy in the hybrid system which in turn leads to the broken $PT$ symmetric phase.

Next, by taking quantum noises into account Eq. (\ref{a9}) can be written into more compact form as $\dot{u}(t)=Au(t)+n(t)$. Here, $u(t)$ is the same $CV$ state vector, $A=-\dot{\iota}\Tilde{H}$ is the drift matrix and $n^{T}(t)=\sqrt{\kappa}X^{in}, \sqrt{\kappa}Y^{in},\sqrt{\gamma}Q^{in},\sqrt{\gamma}P^{in}$ is the matrix of corresponding noises. The input noise quadratures used in $n^{T}(t)$ are defined as: $X^{in}\equiv(\Tilde{c}^{in}+{\Tilde{c}^{in^{\dagger}}})/\sqrt{2}$, $Y^{in}\equiv(\Tilde{c}^{in}-{\Tilde{c}^{in^{\dagger}}})/\dot{\iota}\sqrt{2}$, $Q^{in}\equiv(\Tilde{b}^{in}+{\Tilde{b}^{in^{\dagger}}})/\sqrt{2}$, and $P^{in}\equiv(\Tilde{b}^{in}-{\Tilde{b}^{in^{\dagger}}})/\dot{\iota}\sqrt{2}$. The solution of this Langevin equation is given as $u(t)=e^{At}u(0)+\int_{0}^{t}dse^{A(t-s)}n(s)$. The possible solutions of the system can be found for negative eigenvalues of $A$.  It follows that only when the system remains in the $PT$ symmetric phase, we can find such solutions.

We also observe that the system preserves its Gaussian properties due to the above-linearized dynamics and zero-mean Gaussian nature of the quantum noises. To completely describe the system, one can use the usual covariance matrix (CM) approach \cite{RevModPhys.77.513}. Let $W_{ij}(t)$ be the CM with each
element defined as 

\begin{equation}\label{a12}
      W_{ij}(t)= \frac{\bra{z}{u_{i}(t)u_{j}(t) + u_{j}(t) u_{i}(t)}\ket{z}}{2}.
\end{equation}
   Here, we consider a squeezed state as  input state, i.e., $ \ket{z} = e^{r(c^\dagger b^\dagger-c b) }\ket{0,0}$, which is
 a two-mode squeezed  vacuum state,
with $r$ being the squeezing parameter. For realization of such squeezed states in a hybrid  system one may look into other striking proposals in Refs. \cite{PhysRevLett.107.123601,Riedinger-2018,Ockeloen-Korppi-2018,PhysRevLett.121.220404}. The  equation of motion as satisfied by the CM is given by:
\begin{equation}\label{a13}
    \dot{W}(t) = AW(t) + W(t)A^T + Z.
\end{equation}
Here, $ Z $ defines the matrix of noising correlation which is equal to $ Z= [ \frac{\gamma}{2} + \kappa (n_{th} + \frac{1}{2} )] $ can be obtained by using the Markovian assumption and $ ({n_{i}(t)n_{j}(t) + n_{j}(t) n_{i}(t)})/{2} = \delta (t-t^{'}) Z_{ij}$.
Eq. (\ref{a13}) is a first-order non-homogeneous differential equation that can be solved analytically and numerically by using proper initial conditions (Please see the appendix for an analytical solution). Here, our main concern is to investigate quantum entanglement.  The formal solution of the Eq. (\ref{a13})  can be written as 
\begin{equation}
    W=\left(
\begin{array}{cccc}
 W_{A} &  W_{AB} \\
 W_{AB}^T & W_{B} \\  
\end{array}
\right),
\end{equation}
where $W_{A}$, $W_{AB} $, $W_{B}$ and $W_{AB}^T $  are the $ 2\times2 $ sub-matrices, respectively, corresponding to the local covariance matrices of the CPW microwave cavity and the NAMR and the non-local correlation between them. One can calculate  the so-called logarithmic negativity $E_N$  to determine the degree of quantum entanglement, such as
    \begin{equation}
 E_{N}= max[0, -ln(2W^-)],  
 \end{equation}
 where $W^- \equiv \frac{1}{\sqrt{2}}{\left[\sum (W)- \sqrt{\sum{(W)}^2- 4\det W}\right]}^{\frac{1}{2}}$ is the smallest simplistic eigenvalue of the partial transpose of the $W$ with $\sum (W)\equiv \det(W_{A}) + \det(W_{B}) -2\det(W_{AB})$. It is well established that the relative phase of the input fields plays important role in  entanglement dynamics. Therefore, we also investigate nonclassicality feature and try to build a correlation between the entanglement and the antibunching. For  the cavity  and mechanical resonator modes $c$ and $b$, respectively, the condition for inter-mode  antibunching is defined as follows \cite{PhysRevA.99.023820}
 \begin{equation}\label{16}
   \mathcal{A}(b,c)= \frac{\langle  b^{\dag} c^{\dag}bc\rangle-\langle  b^{\dag}b \rangle \langle  c^{\dag}c \rangle}{\langle  c^{\dag}c \rangle \langle  b^{\dag} b \rangle} .
 \end{equation}
  \begin{figure}
    \centering
    \includegraphics[height=6cm,width=9cm]{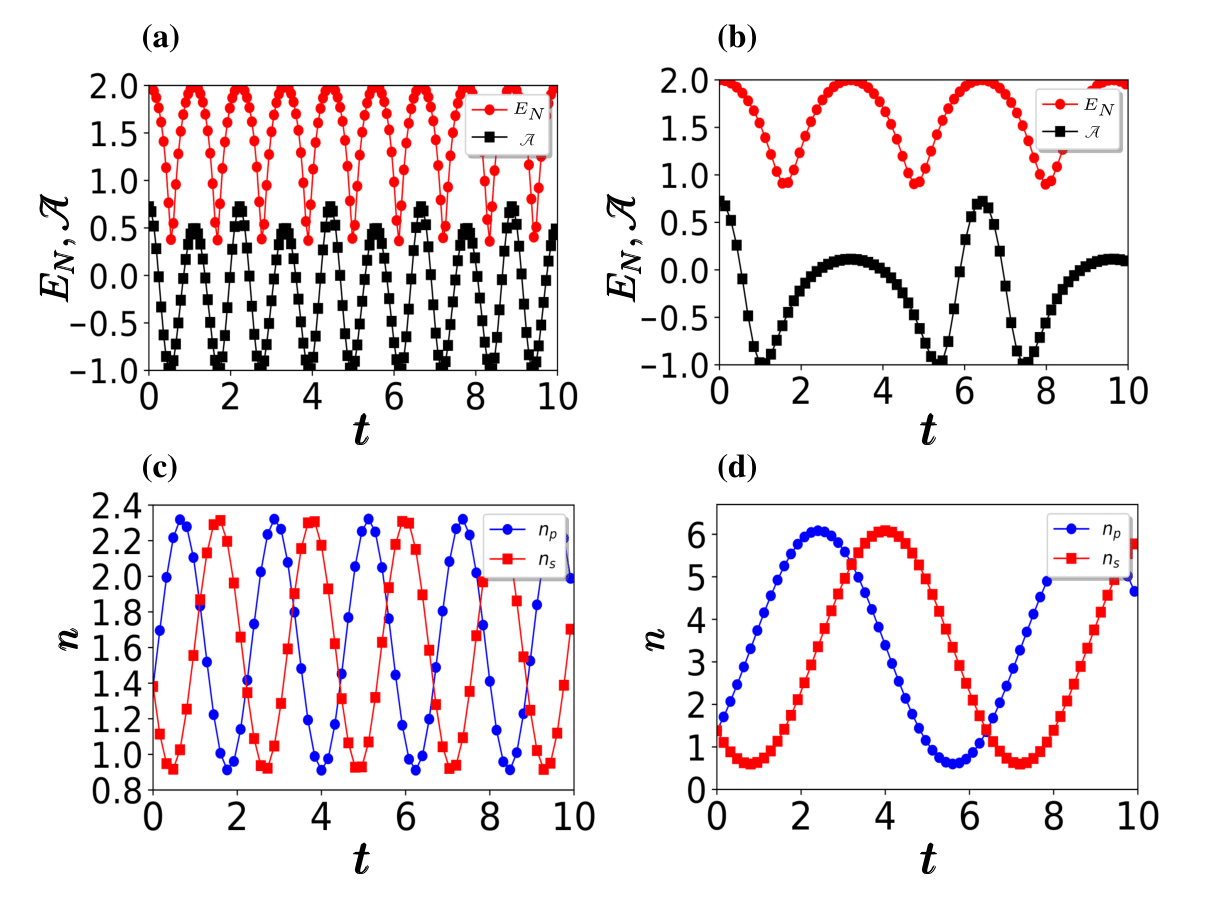}
\caption{For $s=1$  (a) Entanglement evolution (line-circles) between the gain and loss resonators with antibunching (line-squares) for $G=1.5$  (b) for  $G=0.7$. Dynamics of photon numbers (line circles) and phonons (line-squares) numbers for $G=1.5$  (c) and $G=0.7$ (d).  }
    \label{Fig-3}
\end{figure}
 Here, the first term on the right-hand side represents the simultaneous detection in the output of two coupled microwave cavities and the NAMR, while the second term defines the product of individual photon and phonons numbers. The time evolution of our system for the balanced gain-loss case is shown in Fig. (\ref{Fig-3}). We have observed that when the system is in $PT$-symmetric phase the entanglement, antibunching, number of photons $n_{p}$, and number of phonons $n_{s}$ oscillate periodically.

This oscillation could be ascribed by the nature of the eigenvalues $\pm\sqrt{G^{2}-G_{c}^{2}}$ of $A$, as obtained for $G>G_{c}$. In Fig. \ref{Fig-3}(a), for $G=1.5$, one can see that the entanglement oscillates rapidly with a smaller period and antibunching which is a non-classical feature oscillates briskly and have a positive value for maximum entanglement. However, when we approach near to the  EP i.e., $G=0.7$, we observe lesser oscillation with a smaller amplitude and longer period. So, in the close vicinity of the EP, the entanglement almost freezes out, which means that a longer period is required to complete one oscillation, as shown in Fig. \ref{Fig-3}(b).  We investigate the acute entanglement happens only when the number of photons $n_{p}$ and phonons $n_{s}$ becomes equal during their oscillations as predicted in Fig. \ref{Fig-3}(c,d).    One can observe that at $G=1.5$, i.e., away from the EP,  $n_{p}(t)$ and $n_{s}(t)$ oscillate very rapidly with very smaller time period. Whereas, in the close vicinity of the EP,  $n_{p}$ and $n_{s}$ oscillate slowly with a longer period. The oscillatory behavior of the curves is attributed to the fact that the elements of the $W$ matrix are sinusoidal. We also noticed that the entanglement becomes minimum when the difference of $n_{p}(t)$ and $n_{s}(t)$ becomes maximum and the entanglement  becomes maximum when change between $n_{p}(t)$ and $n_{s}(t)$ becomes minimum. These results show that a direct connection between the number of photon/phonons, antibunching, and entanglement. 

\begin{figure}
    \centering
   \includegraphics[height=3.8cm,width=8.8cm]{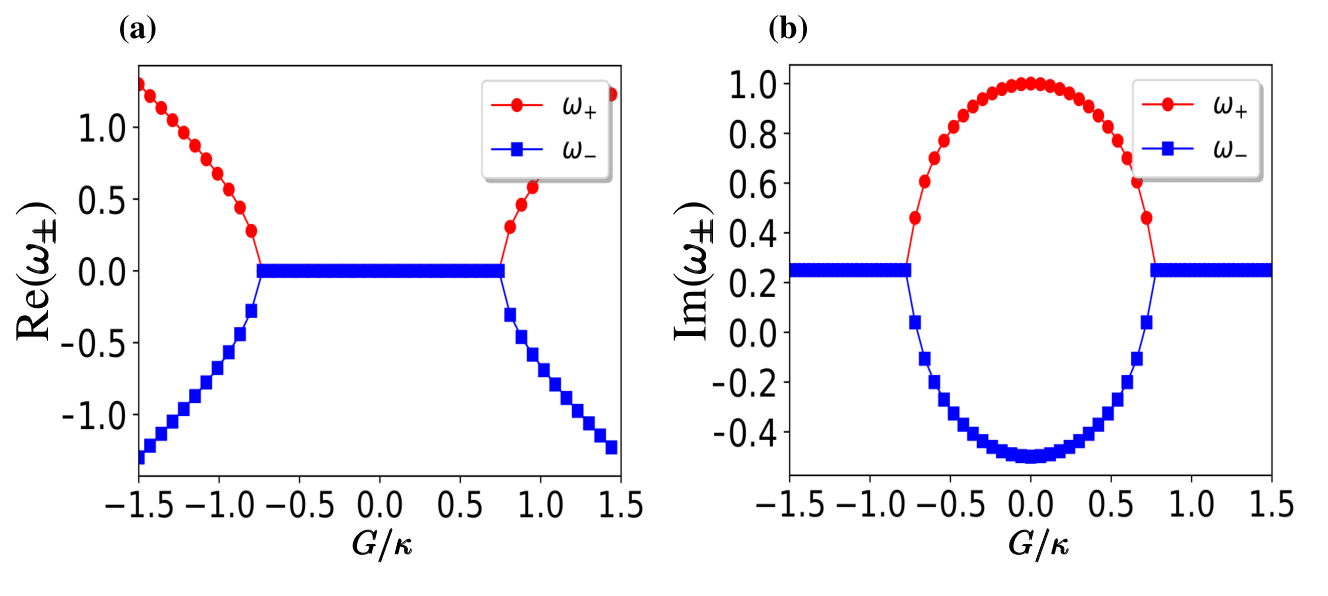} 
\caption{ For the case of unbalanced gain-loss i.e., $\gamma \neq \kappa $ (a) the real and (b) the imaginary  parts the eigenfrequencies $\omega_{\pm}$  are plotted as a function of dimensionless coupling strength $G/\kappa $, here the value of $s=2 $.}
    \label{Fig-4}
\end{figure}

\begin{figure}
    \centering
    \includegraphics[height=5.5cm,width=8cm]{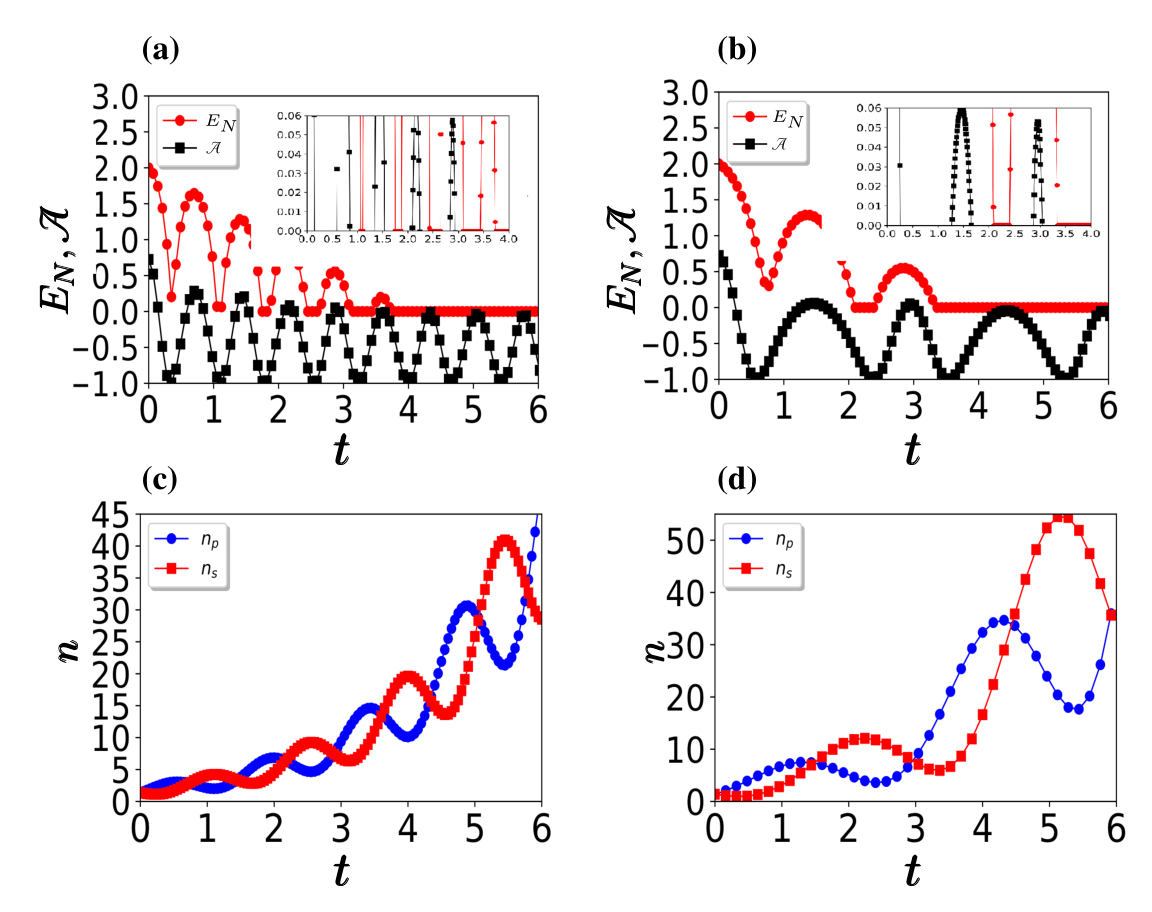}
\caption{For  $s=2$ (a) Entanglement (line-circles) and antibunching (line-squares) evolution for the hybrid system for $G=2.3$ (b) for  $G= 1.3$. In (a) and (b), inlet is plotted to emphasize the positive values of antibunching.  Dynamics of photon numbers (line-circles) and phonons numbers (line-squares) for $G=2.3$  (c) and $G=1.3$ (d).}
    \label{Fig-5}
\end{figure} 
In this subsection, we consider the general situation when the mechanical gain and the cavity loss are not balanced, i.e., $s \neq 1$, with $s$ being 
the gain-loss ratio. For the gain-loss ratio  $s=2$, i.e., $\gamma=2\kappa $ means that in our system mechanical gain is larger as compared to the cavity-loss. In Fig. \ref{Fig-4}(a, b), we plot the eigenfrequencies of the system, which looks identical to Fig. (\ref{Fig-2}), however, the imaginary part of the eigenfrequency is shifted at $0.25$ from zero, this shifted eigenfrequency responsible for the gain in the overall system. In another way, we can say that in an unbalanced gain-loss case eigenfrequencies $\omega_\pm$ are not purely real and the system will always be in broken $PT$ symmetric phase, as shown in Fig. \ref{Fig-4}(a, b). As compared to the balanced gain-loss rate, we observe that with the larger gain, the critical point $G_{c}$ has been shifted to the right and a stronger coupling strength $G$ is required in this situation to enter into a so-called exceptional point.   Moreover, we also study the dynamics of our system for the unbalanced gain-loss case. In Fig. \ref{Fig-5}(a), we show the entanglement evolution for $s=2$. One can see that for unbalanced gain-loss cases, oscillations decrease with time and eventually decays to zero a typical entanglement sudden death-like behavior, which means that unbalanced gain-loss rate adds noise into our system due to which entanglement ceases to oscillate and eventually dies out. However, a notable feature is the delayed entanglement death which can be achieved by pushing the system away from  the EP i.e., $G=0.75$, as predicted in Fig. \ref{Fig-5}(b). We also investigate the antibunching phenomenon for this special scenario which shows a positive magnitude for the maximum entanglement. We note that the non-classical feature i.e., antibunching phenomenon can not be measured for weak entanglement. In the sub-Fig. \ref{Fig-5}(a,b) we plot the zoom out of the antibunching phenomenon, to show the positive magnitude of the antibunching phenomena.  In this case, we also note that when the difference of $n_{p}(t)$ and $n_{s}(t)$ becomes maximum, the entanglement grows and vise versa.   We realize that for the unbalanced gain-loss case both $n_{p}$ and $n_{s}$ grow together, waning the distinction between gain and loss hybrid system.  Since there is more gain in the system, therefore the  $n_{p}$ and $n_{s}$ increase and it is due to external pumping into the system. We also find out that the number of photons and phonons grows exponentially which leads to a classical system therefore the entanglement dies out even though the difference between $n_{p}(t)$ and $n_{s}(t)$ becomes zero.

\begin{figure}
    \centering
    \includegraphics[height=5.5cm,width=8cm]{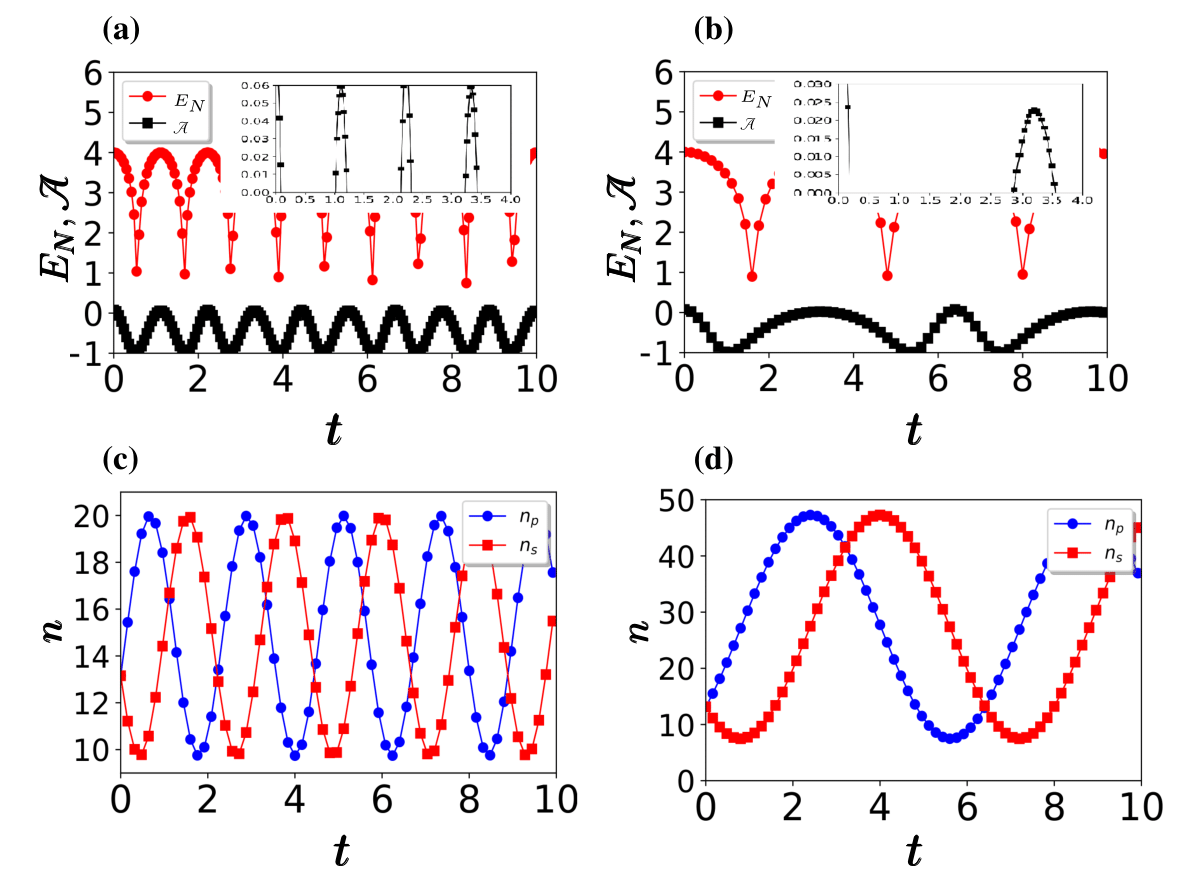}
\caption{Entanglement (line-circles) and antibunching (line-squares) evolution of hybrid electromechanical system for $G=1.5$ (a) and for $G=0.7$ (b) for  $s=1$ and $r=2$. In (a) and (b), inlet is plotted to emphasize the positive values of antibunching. Dynamics  of number of photons (line-circles) and phonons (line-squares)  for $G=1.5$  (c) and $G=0.7$ (d). }
    \label{Fig-6}
\end{figure} 

\section{Altering the entanglement by squeezing parameter}\label{Sec-5}
\begin{figure}
    \centering
    \includegraphics[height=6.0cm,width=8.5cm]{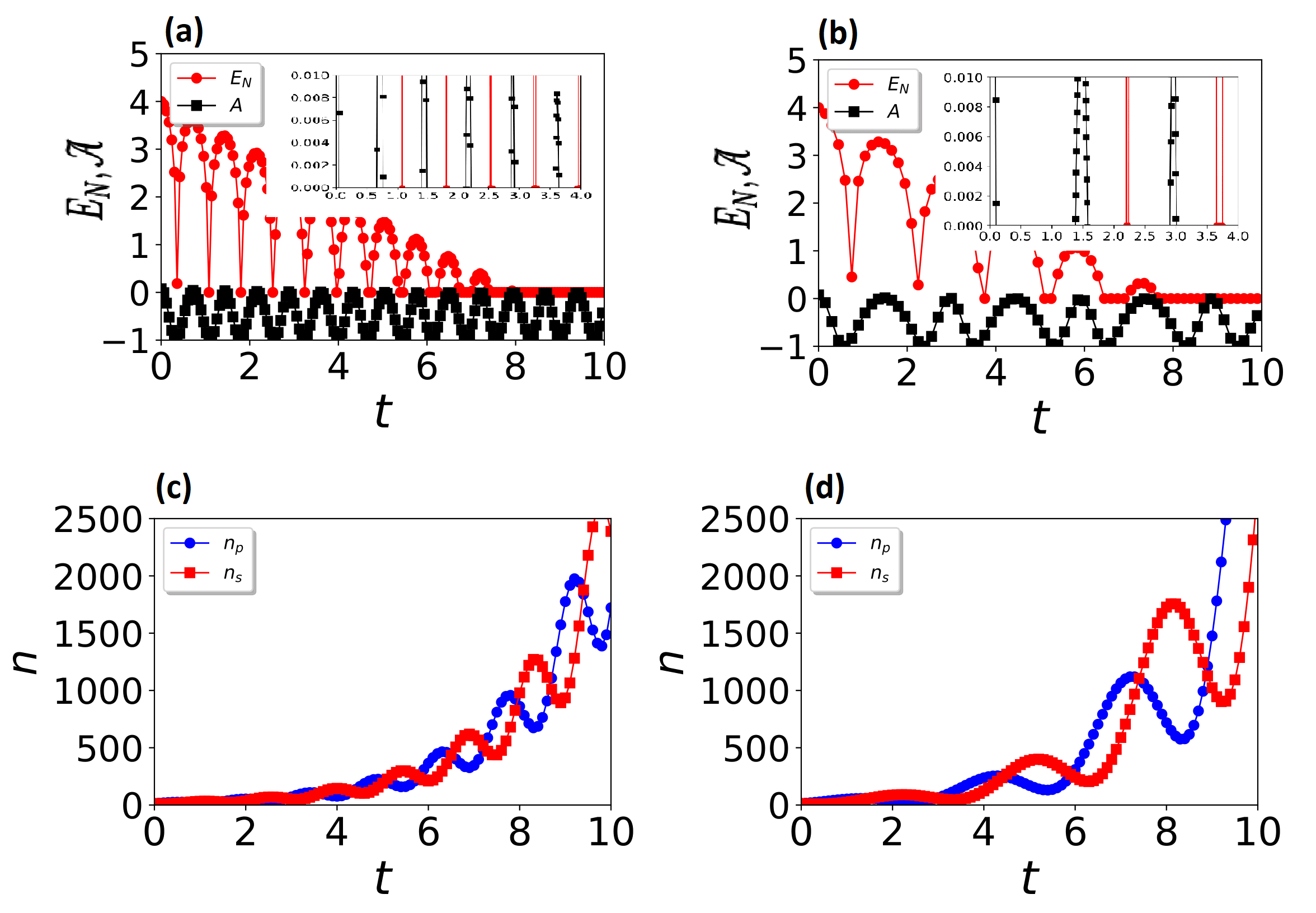}
\caption{Entanglement (line-circles) and antibunching (line-squares) evolution of hybrid electromechanical system for $G=2.3$ (a) and for  $G=1.3$ (b) for  $s=2$ and $r=2$. In (a) and (b), inlet is plotted to emphasize the positive values of antibunching. Dynamics  of number of photons (line-circles)  and phonons (line-squares)  for $G=2.3$  (c) and $G=1.3$ (d).   }
     \label{Fig-7}
\end{figure}

\begin{figure}
    \centering
    \includegraphics[height=4.0cm,width=8.5cm]{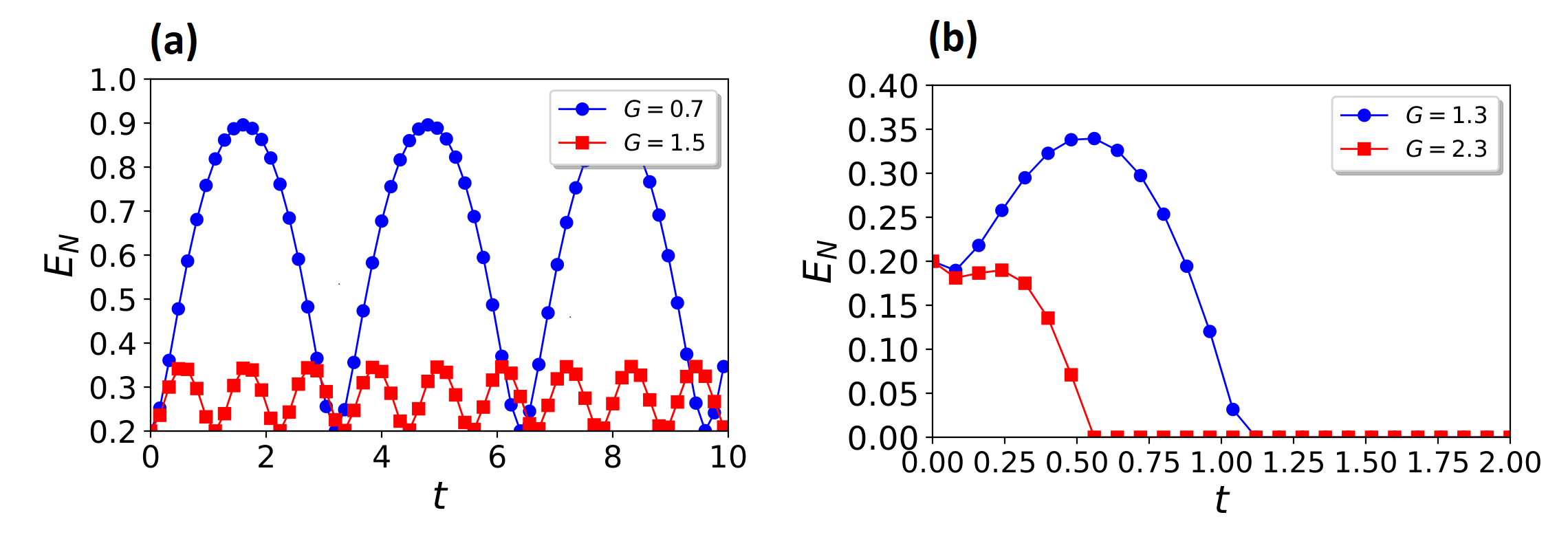}
\caption{Entanglement evolution of  hybrid electromechanical system for  balanced $s=1$ (a) and for unbalance $s=2$ (b)    system,  when the squeezing parameter is defined as $  r= 0.1 $} \label{Fig-8}
\end{figure}

In this section, we investigate the effect of squeezing parameter  $r$ on quantum engagement dynamics for balanced and unbalanced hybrid $PT$ electromechanical systems. 
For the \textbf{balanced} $s=1$ cavity loss and mechanical gain, and for the squeezing parameter $r=2$, the logarithmic negativity oscillates between $[0,4]$ as shown in Fig. \ref{Fig-6}(a-b). The initial average number of photons is increased ten times as predicted in Fig. \ref{Fig-6}(c-d), in comparison to the scenario when squeezing parameter  $r=1$ as represented in Fig. \ref{Fig-2}(c-d). We note that the change in the squeezing parameter has no big impact on the oscillation of entanglement, antibunching, and the number of photon/phonons other than their magnitude as depicted in Fig. \ref{Fig-2}(a-d) and  \ref{Fig-6}(a-d). 
For the \textbf{unbalanced} $s=2$ cavity loss and mechanical gain, and for the squeezing parameter $r=2$, we note that the  so-called sudden death of the entanglement time delay is increased predicted in Fig. \ref{Fig-7}(a-b) as compared to the Fig. \ref{Fig-5}(a-b). It means that the strong squeezing prolongs the sudden death of the entanglement and reduces the influence of quantum noise on entanglement. The strong squeezing also increases the number of photons and phonons as depicted in Fig. \ref{Fig-7}(c-d).  For the weak squeezing $r=0.1$, we note that the dynamics of the entanglement oscillate periodically. Additionally, we also find out that the amplitude of the entanglement dynamics increases near the exceptional point for the balanced gain and loss system as shown in Fig.  \ref{Fig-8}(a). For the unbalanced gain and loss system, we note that the significant delay in entanglement sudden death can be realized by pushing the electromechanical quantum system towards an exceptional point as illustrated in Fig. \ref{Fig-8}(b).  

\section{Conclusion and Summary} \label{Sec-6}
In this study, we studied a  gain-loss hybrid electromechanical system and emphasized the conditions necessary for exhibiting parity-time ($PT$) invariance. To realize a strong and tunable coupling between a Coplanar-Waveguide (CPW) microwave cavity and a nanomechanical resonator (NAMR) we purposed coupling of this hybrid system through a superconducting Transmon qubit.  We examine the hybrid electromechanical system as a non-hermitian Hamiltonian with balanced/unbalanced gain and loss. The gain in our system is introduced in the NAMR and the loss of the system is defined for the CPW when gain and loss are balanced in the hybrid system the energy spectrum gets a real spectrum after the exceptional point. However, when the gain and loss do not equal, then the energy spectrum has some constant imaginary value that defines the gain in the system.  With this setting, we examined the entanglement, antibunching,  average photon, and phonon number with the initial squeezed state. We found out that the time of entanglement sudden death can be delayed near the exceptional points.

We also noticed that the entanglement gets maximum when the difference between the photons and phonons numbers becomes minimum. We further examined a non-classical feature like antibunching for different balanced and unbalanced scenarios. We found out that the entanglement gets maximum when the antibunching gets positive.   Later, we showed that for unbalanced gain-loss cases, oscillations decrease with time and eventually decay to zero a typical entanglement sudden death-like behavior, which means that an unbalanced gain-loss rate adds noise into our system due to which entanglement ceases to oscillate and eventually dies out. In the end, we investigated the dependence of the entanglement dynamics on the initial squeezing parameter, we analyzed that the initial squeezing parameter is not affecting the dynamics of the entanglement for the balanced gain and loss scenario. However, for unbalanced gain and loss, the sudden death of entanglement can be delayed by increasing the initial squeezing parameter. For the weak initial squeezing parameter,  the significant delay in entanglement sudden death can be achieved even in a noisy environment. The present study helps to understand the role of  $PT$-symmetry in the dynamics of the entanglement and in facilitating the preservation of the entanglement for a longer period. With this study, we build a relation between entanglement, antibunching, and the photons and phonons numbers in the physical systems, which are very much relevant in the field of quantum optics and quantum information processing.
 
 Finally, we give a brief description of the experimental prospect of our proposed system. In our system, an external dc voltage source was applied to the NAMR, which established the coupling between the NAMR and the qubit. In a recent experiment \cite{Pirkkalainen-2015}, the coupling between the NAMR and the qubit can be varied between 80 MHz to 160 MHz, by adjusting the external voltage (${V_{NR}=5-10 V}$). Thus, in our system, a strong and tunable coupling between the NAMR and the CPW microwave cavity mediated via Transmon qubit can be achieved. With this setup, the transfer of quantum information between the NAMR and CPW microwave cavity can be accomplished. On the other hand, it has been reported in many experimental studies that the mechanical gain can be achieved via directly driving the mechanical modes including Josephson Phase qubit \cite{Connell-2010}, piezoelectric pump \cite{Xiong-2013},  microwave electrical driving \cite{Bochmann-2013},  and by phonon lasing method \cite{Bahl-2012}. Hence, by considering both the cavity gain and the mechanical loss, the transition from the broken $\bm{PT}$-symmetry phase to the unbroken $\bm{PT}$-symmetry phase can be obtained in our system.  

\section{Acknowledgment}
Jameel Hussain gratefully acknowledges support from the COMSATS University Islamabad for providing him a workspace. 

\bibliographystyle{apsrev4-1}


%

\end{document}